\documentclass[aps,prl,preprint,unsortedaddress,floatfix]{revtex4}
\usepackage{amsthm}
\usepackage{amsfonts}
\usepackage{siunitx}
\usepackage{amsmath}
\usepackage{amssymb}
\usepackage{graphicx}
\usepackage{verbatim}
\usepackage[colorlinks]{hyperref}
\usepackage{tikz}
\usepackage{pgfplots}
\usepackage{braket}
\usepackage{xcolor}
\usepackage{color}
\usepackage[normalem]{ulem}

\definecolor{linkcolor}{RGB}{0,83,166}
\hypersetup{
  colorlinks = true,
  allcolors = {linkcolor}
}

\begin{document}

\title{Kagome qubit ice}

\author{Alejandro Lopez-Bezanilla$^{1}$}
\author{Jack Raymond$^{2}$}
\author{Kelly Boothby$^{2}$}
\author{Juan Carrasquilla$^{3,4}$}
\author{Cristiano Nisoli$^{1}$}
\email[]{cristiano@lanl.gov}
\thanks{These two authors contributed equally}
\author{Andrew D. King$^{2}$}
\email[]{aking@dwavesys.com}
\thanks{These two authors contributed equally}

\affiliation{$^1$Theoretical Division, Los Alamos National Laboratory, Los Alamos, New Mexico 87545, USA}
\affiliation{$^2$D-Wave Systems, Burnaby, British Columbia, Canada, V5G 4M9, Canada}
\affiliation{$^3$Vector Institute, University of Toronto, Toronto, Ontario, M5G 1M1, Canada}
\affiliation{$^4$Department of Physics and Astronomy, University of Waterloo, Waterloo, Ontario, N2L 3G1, Canada}

\date{\today}


\begin{abstract}
Topological phases of spin liquids with constrained disorder can host a kinetics of fractionalized excitations. However, spin-liquid phases with distinct kinetic regimes have proven difficult to observe experimentally. Here we present a realization of kagome spin ice in the superconducting qubits of a quantum annealer, and use it to demonstrate a field-induced kinetic crossover between spin-liquid phases. Employing fine control over local magnetic fields, we show evidence of both the Ice-I phase and an unconventional field-induced Ice-II phase. In the latter, a charge-ordered yet spin-disordered topological phase, the kinetics proceeds via pair creation and annihilation of strongly correlated, charge conserving, fractionalized excitations. As these kinetic regimes have resisted characterization in other artificial spin ice realizations, our results demonstrate the utility of quantum-driven kinetics in advancing the study of topological phases of spin liquids.\end{abstract}

\maketitle
\newpage

\section{Introduction}

Dynamics in crystals typically proceeds via motion of topological defects such as dislocation gliding~\cite{nelson2002defects}. One might expect the kinetics of disordered systems to be naturally free. But in spin liquids, where disorder is present but constrained, kinetics often also proceeds through defects or excitations are endowed with a conserved topological charge \cite{henley2011classical}. For instance, frustrated spin systems, such as pyrochlore~\cite{matsuhira2002new,bramwell2020history}  or square~\cite{perrin2016extensive,farhan2019emergent,King2021} spin ices, remain disordered at low temperature, leading to a Pauling residual entropy, and their disorder is constrained by the so-called ice rule~\cite{pauling1935structure}. There, kinetics consists of creation/annihilation and walks of localized violations of the ice rule, in the form of emergent magnetic monopoles~\cite{Castelnovo2008} that conserve a topological charge. 

Among spin ices, the  kagome ice model~\cite{shastry1981exact,wills2002model,moessner2003theory,Moller2009,Chern2011,libal2017,balents2010spin,raban2019multiple,lhotel2020fragmentation} has been widely studied because it  mimics a remarkable variety of natural and artificial systems, from rare-earth pyrochlores~\cite{matsuhira2002new}, to nanomagnetic fabrications~\cite{qi2008direct}, gravitationally trapped colloids~\cite{libal2017}, and many other systems ~\cite{libal2009,wang2018switchable,xue2018tunable,Mellado2012,duzgun2021skyrmion,zhao2020realization,hua2021highly,meeussen2020topological,pisanty2020topological}.  Kagome spin ice can in principle manifest various unusual phases~\cite{Moller2009,Chern2011,libal2017}, but the large  energy scales of artificial implementations pose an experimental challenge; thorough measurements of these phases and the physical conditions driving the phase-to-phase transition are scarce. 

Here we present a kagome qubit ice realized in a superconducting quantum annealer.  Using this experimental platform, we study its field-induced spin-liquid phases and quantum-activated kinetics. We experimentally establish that topological constraints affecting the dynamics proceeds via charge-conserving fractionalized excitations. Using thousands of programmable external magnetic fields, we detune the system from its more common ice-rule-obeying ``Ice-I'' phase into a field-induced ``\mbox{Ice-II}'' phase, which exhibits charge order while remaining spin-disordered.

\section{Results}

\subsection{Kagome spin ice}

Kagome spin ice consists of magnetic dipoles as classical binary Ising spins arranged along the edges of a hexagonal lattice and therefore on the sites of a kagome lattice. They point from one triangular ``ice vertex'' (kagome plaquette) to another (Fig.~1a).  We can thus introduce the notion of a magnetic charge for a vertex, defined as the number of spins pointing toward the vertex minus those pointing away from it. Because of the odd coordination, a vertex can host only nonzero, odd charges $q=-3, -1, 1, 3$ (Fig.~1b).

The simplest magnetic kagome model includes interactions only among spins impinging on the same vertex. Since not all pairs of spins at a vertex can simultaneously assume an energy-minimizing head-to-tail configuration, the system is  {\it frustrated}.  The ground state is therefore an extensively degenerate ensemble of disordered spins obeying the (pseudo-) ice rule: frustration is minimized when each vertex has two spins pointing in and one pointing out, or vice-versa. This {\it ice manifold} is often called the {\it Ice-I phase}, and can be thought as a spin liquid forming an overall neutral plasma of disordered $\pm1$ magnetic charges. In the {\em \mbox{Ice-II}} phase \cite{Moller2009,Chern2011,macdonald2011classical}, disordered spins still still obey the ice rule but charges are ordered in an ionic lattice~\cite{Zhang2013,drisko2015fepd,levis2013thermal,anghinolfi2015thermodynamic}.

\subsection{Kagome qubit ice}

In this work, we realize kagome spin ice in a quantum annealer. Its superconducting flux qubits are described by the transverse field Ising Hamiltonian
\begin{equation}\label{eq:tfim}
  H_Q = -\Gamma\sum_{i} \hat \sigma_i^x + \mathcal J\big(\sum_ih_i\hat \sigma_i^z +\sum_{ij}J_{ij}\hat \sigma_i^z\hat \sigma_j^z \big),
\end{equation}
where $\hat \sigma^x$ and $\hat \sigma^z$ are Pauli matrices on the qubits, $\mathcal J$ is an energy prefactor on the classical Ising Hamiltonian, $h_i$ are per-qubit programmable longitudinal fields~\cite{harris2010experimental}, and $J_{ij}$ are programmable two-qubit couplers. $\Gamma$ is a transverse field entangling the Pauli matrices and thus controls quantum fluctuations.

Kagome spin ice can be mapped to a classical Ising model \cite{zhang2012perpendicular}, and therefore to the Hamiltonian of Eq.~(\ref{eq:tfim}).  Consider alternating A and B vertices pointing up ($\vartriangle$) and down ($\triangledown$) respectively  in Fig.~1a.  We assign an Ising spin value $s_i=+1$ if it points into the A vertex, and $s_i=-1$ if it points into the B vertex (Fig.~1b). (Compare with the   Then, standard kagome ice  corresponds to the Hamiltonian
\begin{equation}\label{eq:ising}
  H_I = J \sum_{\langle i,j\rangle}s_is_j + \sum_ih_is_i
\end{equation}
where each nearest-neighbor spin is coupled antiferromagnetically. 

We then embed the kagome lattice in the graph of available two-body couplers, as shown in Fig.~1c, by modifying an embedding of a $\mathbb Z_2$ lattice gauge theory into the transverse-field Ising model~\cite{Chamon2020}.  Each kagome site is represented by a ferromagnetic three-qubit chain, and nearest-neighbor chains are coupled antiferromagnetically with two physical couplers. (Three qubits are needed for each kagome lattice site because it is not possible to directly couple two arbitrarily-chosen qubits.) We use $h$ and $J$ (with no index) to denote the total field on a three-qubit chain and the total coupling between two neighboring chains, respectively, obtaining the kagome qubit ice (KQI)  Hamiltonian
\begin{equation}\label{eq:kqi}
  H_{\textit{KQI}} =  -\tilde\Gamma\sum_{i}\tilde\sigma_i^x + \mathcal J\big(h \sum_i \tilde\sigma_i^z +J\sum_{\langle i,j\rangle}\tilde\sigma_i^z\tilde\sigma_j^z \big),
\end{equation}
where $\tilde\Gamma= \Gamma^3/J_{FM}^2$ is an effective transverse field on the three-qubit chains for a ferromagnetic chain coupling $J_{FM}$~\cite{King2021prxq}, $\tilde\sigma_i$ denotes a logical moment represented by a three-qubit chain, and indices $i$ and $j$ are also over three-qubit chains, rather than individual qubits.

\subsection{Phases}

When $\Gamma=\tilde\Gamma=0$ and $h=0$, the extensively degenerate ground state manifold of $H_{\textit{KQI}}$ corresponds to that of $H_I$, which is the commonly seen Ice-I phase~\cite{qi2008direct}.  But we can go beyond this regime.  In nanoscopic realizations, another phase of lower entropy is possible~\cite{Rougemaille2011,Zhang2013,drisko2015fepd,anghinolfi2015thermodynamic}. In such systems, it is driven by the long range nature of the dipolar interactions~\cite{Moller2009,Chern2011}.  It still has disordered ice-rule obeying spins, but with charges ordered in an ionic lattice where $A$ and $B$ vertices have opposite charge. {While the spins remain disordered, though at lower entropy~\cite{Moller2009,Lammert2010,Chern2011}, their disorder is  topologically constrained: it can be mapped to a dimer cover model~\cite{moessner2003theory,Lammert2010} and considered a case of {\it classical topological order}~\cite{macdonald2011classical,henley2011classical,lamberty2013classical,lao2018classical}}. This is often called the {\em \mbox{Ice-II}} phase, and its topological nature should show a topologically protected kinetics. (Note also that Ice-II can also be considered a broken symmetry phase with unsaturated order parameter in the context of magnetic fragmentation~\cite{lhotel2020fragmentation,canals2016fragmentation,brooks2014magnetic}.)

Indeed, the kinetics in the Ice-I phase is {\it not gapped}: It is possible to flip a single spin---or indeed an extensive number of single spins---without violating the ice rule and thus without creating an excitation (see also Supplementary Informations). Thus, the system can kinetically explore the phase {\it from within} the {local low energy} manifold. 

Instead, in the \mbox{Ice-II} phase any individual spin flip disrupts the charge balance, thus creating an excitation.
Therefore~\cite{henley2011classical}  the kinetics of the  \mbox{Ice-II} phase must proceed either via pair creation, motion, and annihilation of gapped excitations, or else via cooperative, ungapped flips of entire loops of head-to-tail spins which do not alter the charge distribution. 
Such kinetics was never probed in previous realizations of kagome ice because the \mbox{Ice-II} phase has proved very hard to reach~\cite{Rougemaille2011,Zhang2013,drisko2015fepd,anghinolfi2015thermodynamic} (see Supplementary Informations).
Fortunately, the quantum annealer offers another route: {we can induce it by the field $h$, acting on $\sigma^z$, and then} we can study field-induced \mbox{Ice-II} kinetics.  

If we define a staggered charge $q_{s}$  on a vertex such that $q_s=-q$ for $A$ vertices and $q_s=q$ for $B$ vertices, then the field $h$ determines the vertex energies $\varepsilon_{-3}$, $\varepsilon_{-1}$, $\varepsilon_{+1}$, $\varepsilon_{+3}$ for vertices with $q_s=-3$, $-1$, $1$, and $3$ respectively, as shown in Fig.~2 (see also SI).

For $0< h/J< 4$, $\varepsilon_{+1}$ has the lowest energy, leading to the charge-ordered, spin-disordered \mbox{Ice-II} phase as the ground state.
Within this window, Fig.~2 shows a regime crossover at $h/J=2$. The lowest excitations are charge-order violations upsetting the ionic crystals of charges when $0<h/J<2$, and ice rule violations when $2<h/J<4$. The two types of excitations are degenerate at $h/J=2$ where the excitation gap is highest.

Then, for $h/J > 4$, the ground state degeneracy vanishes, replaced by an ordered   state in which all $A$ and $B$ vertices have charge $-3$ and $3$, respectively.

To estimate pseudo-equilibrium properties of the kagome qubit ice in these different phases, we begin with a random spin state and repeatedly expose the system to quantum fluctuations as described by Eq.~(\ref{eq:tfim}), by cycling the transverse field $\Gamma$ on and off.  An appropriate magnitude of transverse field drives the kinetics of this kagome qubit ice without erasing the state memory, as previously demonstrated in square ice \cite{King2021}.  After each exposure, we read out a classical spin state.  This leads to a sequence of states amenable to statistics (see SI).

Fig.~3 summarizes experimental results for varying $h/J$.  Fig.~3a shows real-space samples, represented as vertex charges, for increasing values of $h$. 
At $h=0$ we see the expected disordered charge plasma of the Ice-I phase.  Increasing $h$ first leads to  ionic ordering of the charge ( the \mbox{Ice-II} phase)  eventually giving way to a polarized state in which the longitudinal field overcomes the ice rule, forming ionic crystals of $\pm3$ charge, and all spins have value $s_i=1$.

Fig.~3b shows the corresponding result in reciprocal space via the Fourier transform of the spins defined as 
   $ S({\bf q}) \propto \sum_{ij} e^{i{\bf q}({\bf r}_i-{\bf r}_j)} \left( \langle s_is_j \rangle - \langle s_i \rangle\langle s_j \rangle\right)$.
Our sign convention for the spins leads to the appearance of peaks only in the \mbox{Ice-II} phase and its proximity, and the formation of pinch points in the topologically protected region with $h/J=2.5$. In Fig.~3c, cuts of the  {Fourier transforms} through the high-symmetry points in the extended Brillouin zone clearly show growing peaks at $K$ {in the proximity of the Ice-II phase}. {These peaks correspond to the expected logarithmic divergence of the dipolar correlations~\cite{moessner2003theory} (see also Fig. 5 in ref~\cite{moessner2003theory}, obtained from a dimer model). They, and the pinch points, follow therefore from the topological properties induced on the phase by the charge ordering}. From an implementation point of view, $S({\bf q})$ reveals a highly symmetric system in which the multi-qubit embedding of kagome spins preserves isotropy. This is an important advance over previous work \cite{King2021}.

Fig.~3d plots the {\it charge order parameter}, defined as one third the average staggered charge of a vertex.  The two two broad plateaus at $\pm 1/3$ correspond to the \mbox{Ice-II} phases.  

Fig.~3e confirms the high ice-rule obedience throughout the Ice-I and \mbox{Ice-II} phases, which breaks down at $h/J>|4|$ where, from Fig.~2, the lowest energy vertex no longer obeys the ice rule.

\subsection{Topologically protected quasi-classical kinetics}

These measurements validate the annealer's effectiveness as an experimental platform for probing phases of the Ising kagome spin ice system near a low-temperature thermal equilibrium.  Because consecutive output states are separated dynamically by a relatively short exposure to a relatively weak transverse field $\Gamma$ (compared to $J$), we can also probe the quasi-classical kinetics.

As mentioned above, in the \mbox{Ice-II} ground state a single spin flip always corresponds to fractionalized excitations, as either violations of the \mbox{Ice-II} charge-order constraint, or violations of the kagome ice rule (Fig.~2). We can define a topological charge (or t-charge) as $q_t= q+1$,  $q_t= q-1$ for $A$ and $B$  vertices respectively. In the \mbox{Ice-II} charge-ordered ground state, the topological charge is zero on all vertices. Instead, excitations of the \mbox{Ice-II} phase are topologically charged. Their t-charge is conserved: flipping a spin creates a pair of fractional excitations of t-charges $\pm2$ and zero net t-charge. Further flips can separate the t-charges, which can then annihilated when meeting other, opposite ones. This situation of paired fractional excitations is very reminiscent of square and pyrochlore ice~\cite{King2021,castelnovo2007topological}, although here the topological charge is not the magnetic charge.

To probe the thermal and quantum-activated kinetics of the Ice-I and \mbox{Ice-II} phases, we compare QA output samples. Between consecutive samples, the qubits are exposed to the a transverse field for $\SI{1}{\micro s}$, and at the same time $\mathcal J$ is dropped.  This protocol is depicted in Fig.~4a.  Since the system is in a thermal bath at $\SI{12}{mK}$, this allows both quantum and thermal fluctuations to drive dynamics \cite{King2021}.

{In agreement with the description above, our results show a kinetics of fractionalized excitations, that can be created and annihilated in pairs of opposite topological charge, and more rarely a kinetics consisting of flips of entire loops of spins---which can always be construed mathematically as creations followed by annihilation of topologically charged pairs.}

Fig.~4b shows two representative samples from each of $h/J = 0.5, 2.5 ,4$, corresponding roughly to the boundaries and the middle of the field-induced \mbox{Ice-II} phase.  Ice-rule and charge-order violations are shown as triangles.  Between the two samples, we highlight the spins that flip during the exposure to fluctuations, as well as the motion of fractional excitations.  

At $h/J=0.5$ the charge order is fragile and we are close to the Ice-I phase. We see many excitations popping up erratically, and they are charge order violations, due to their small energy cost (Fig.~2).  

At $h/J=2.5$ we see far fewer excitations, and the kinetics consists of their wandering. We also see flipping of  closed loops of spins. One fractional excitation escapes off the boundary, one appears from the boundary, and one moves to another location through a chain of flipped spins. This picture is consistent with the large energy gap shown in Fig.~2, which suppresses pair creation of excitations. 

At $h/J=4$, we again see a regime in which excitations can appear at low cost; these cheap excitations are now ice-rule violations, in contrast to the charge-order violations seen near the Ice-I phase, consistent with the energetics (see Fig.~2).

To quantify the creation/annihilation and motion of fractional excitations, we consider the subgraph of the honeycomb lattice whose edges correspond to flipped spins (Fig.~4c--d) between consecutive states. We measure the degrees (valencies) of honeycomb sites in this graph.  A closed loop of flipped spins results in only degree-two honeycomb sites.  Conversely, an open chain of flipped spins will have degree two in the interior, and degree one on the ends.  This can involve the motion of a fractional excitation, with or without creation/annihilation.  In general, degree-two spins correspond to motion of excitations, while degree-one spins correspond to creation/annihilation.  

Fig.~4e shows that the system is overall most active around $h/J=0$ and $h/J=4$, which corresponds to points of degeneracy (see Fig.~2) where excitations are cheapest.  The plot of the relative frequency of excitation motion over pair creation/annihilation shows a maximum around $h/J=2$, the point of maximum gap: where excitations are most expensive, kinetics consists mostly of their random walk, much like monopoles in square or pyrochlore spin ice. 

The non-monotonicity of the curves in Fig.~3e shows that in kagome qubit ice, by tuning the gap of the phase, the topological protection of the kinetics can be controlled, from a hard to distinguish soup of excitations at $h/J=0,4$, to a clear picture of creation/annihilation and motion of fractionalized excitations around the value $h/J=2$.

\section{Discussion}

We have realized kagome qubit spin ice in 2742 superconducting flux qubits of a quantum annealing processor and explored its field-induced spin-liquid ice phases.  We have studied the quantum-activated, topologically protected kinetics of the \mbox{Ice-II} phase and shown that it proceeds via creation/annihilation and propagation of charge-conserving fractionalized excitations.   We emphasize that quantum fluctuations are used here only to drive kinetics, but can be employed in the future to study entangled states. Furthermore, the kagome antiferromagnet in a transverse field $\Gamma$ has a rich ground-state phase diagram \cite{Moessner2000} arising from high-order perturbations in $\Gamma$, which may be probed in future work. Our results demonstrate that quantum annealers are capable of implementing exotic programmable phases of frustrated spin sliquids, whose gap and topologically-protected kinetic regimes can be finely tuned.

\clearpage
\begin{figure*}
  \includegraphics[width=1 \linewidth]{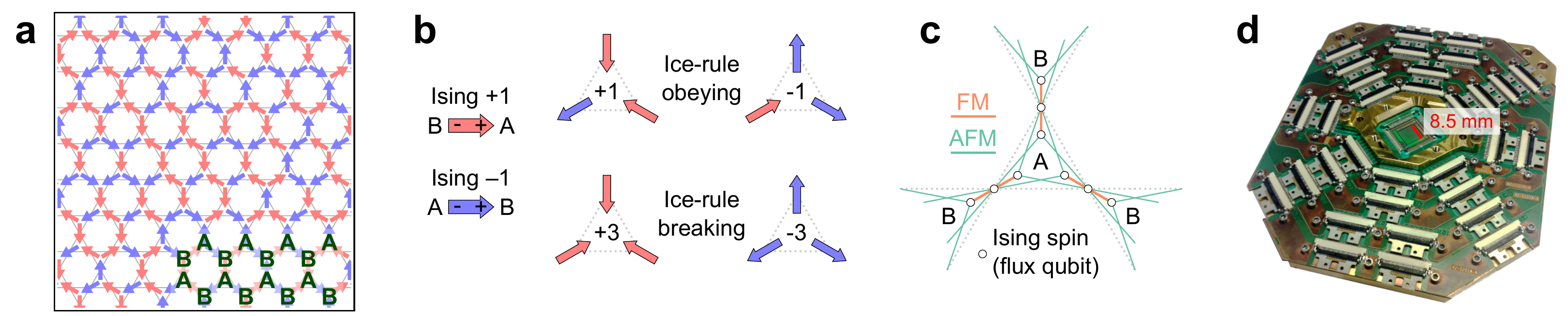}
  \caption{{\bf Kagome qubit ice.}  {\bf A}, Kagome spin ice consists of magnetic dipoles on the edges of a hexagonal lattice, which point in or out of triangular plaquettes (vertices) of the dual kagome lattice (gray lines). {\bf B}, Each vertex in a given configuration has a nonzero charge: $\pm 1$-charged vertices satisfy the kagome ice rule; $\pm 3$-charged vertices do not.    Denoting triangles pointing up and down by A and B respectively, one can map dipoles to Ising spins according to whether not the dipole points into an A triangle.  
    {\bf C},  In the kagome qubit ice, each kagome site is realized using a ferromagnetically-coupled three-qubit chain.  Sites impinging on the same triangular ice vertex are coupled antiferromagnetically, leading to geometric frustration.
  {\bf D},  Optical image of the superconducting quantum annealing processor in a sample holder.  $2742$ qubits are used to realize a $913$-spin kagome ice.}\label{fig:1}
\end{figure*}

\clearpage
\begin{figure}[t!]
  \includegraphics[width=8.5cm]{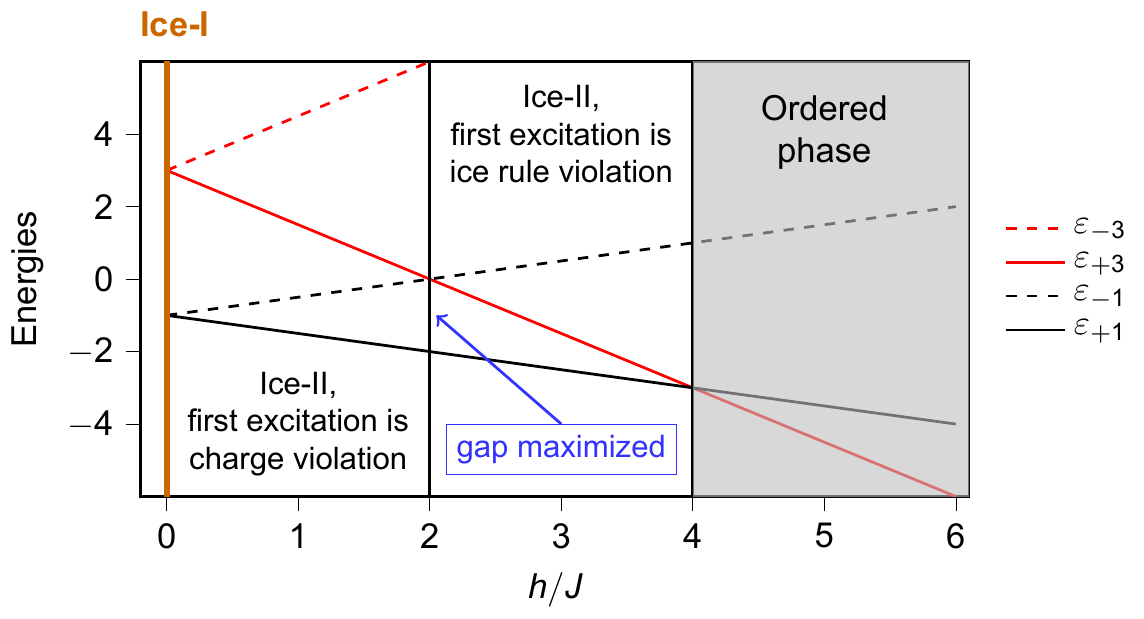}
  \caption{{\bf Ice vertex energies} (normalized to $J$).  In the Ice-I phase ($h=0$), the ice-rule vertex states $\varepsilon_{+1}$ and $\varepsilon_{-1}$ are degenerate.  Detuning $h$ leads energetic preference towards the (staggered) $+1$-charged configurations.  Within the ice region $0\leq h/J\leq 4$, the energy gap is maximized at $h/J=2$, where charge-imbalance excitations are degenerate with ice-rule excitations.}\label{fig:2}
\end{figure}

\clearpage
\begin{figure*}[t!]
  \includegraphics[width=1 \linewidth]{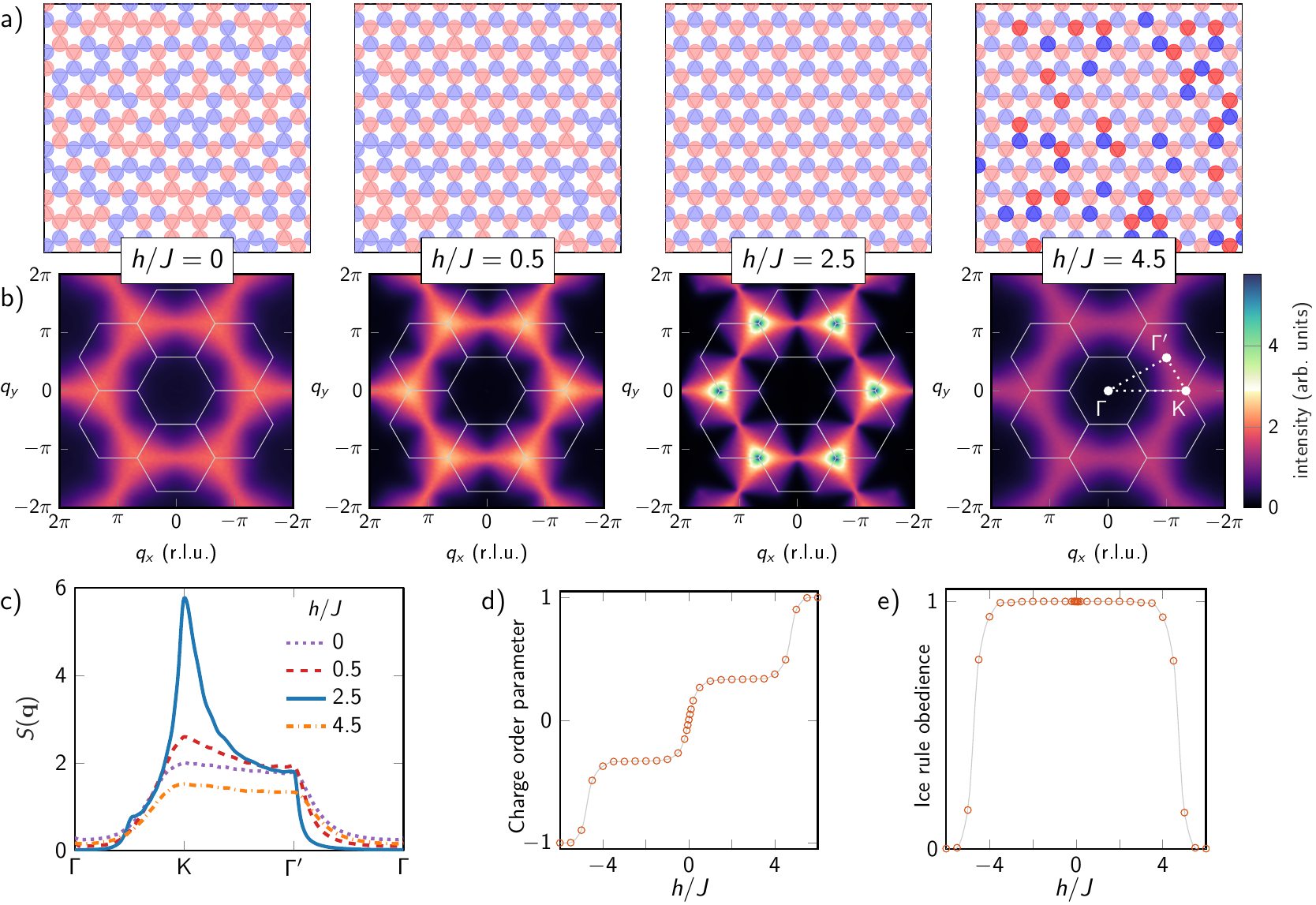}
  \caption{{\bf Field-induced charge phases and qubit ice structure.} {\bf A}, Charge states for varying external field $h$.  At $h=0$, vertices have no energetic preference between $-1$ and $+1$ charge (light blue and red respectively), leading to charge disorder.  As $h$ increases, A and B vertices energetically favor $-1$ and $+1$ charge respectively, leading to long-range order in the staggered charge.  Eventually $h$ polarizes the sites, leading to a preponderance of $-3$ and $+3$ charged vertices (dark blue and red respectively).  {\bf B}, {Fourier transforms} $S({\bf q})$ calculated from QA experimental output, with Brillouin zone in gray.  {\bf C}, Cuts of $S(\mathbf q)$ for varying $h/J$ through high-symmetry points $\Gamma$, $K$, and $\Gamma'$ (shown in {\bf b}) show the effect of the longitudinal field on peak height at $K$ and pinch-point width at $\Gamma'$.  {\bf D}, Charge order parameter.  {\bf E}, Proportion of vertices obeying the ice rule.}\label{fig:3}
\end{figure*}

\clearpage

\begin{figure*}[htp]
  \caption{{\bf Kinetics and field-induced topological protection.} {\bf A}, Within the quantum annealer, the kinetics is driven by a reverse anneal protocol wherein the qubits (Eq.~(\ref{eq:tfim})) are exposed to quantum fluctuations ($\Gamma$) and thermal fluctuations $(T/\mathcal J)$ for a duration of $\SI{1}{\micro s}$ between projected classical output states.
    {\bf B}, Quantum annealer output samples. For each $h/J$, two consecutive states are shown, along with the spin-flip difference between them.  For the states, up and down triangles  denote charge-order violations and ice-rule violations respectively.  For the spin-flip differences, crosses denote excitations in state $i$, and circles denote excitations in state $i+1$.  For $h/J=2.5$, near the middle of the field-induced \mbox{Ice-II} phase, the two excitation types are nearly degenerate, with a large gap.  Thus excitations are suppressed although they can move freely.  {\bf C--D}, Flipped spins ({\bf c}) form a subgraph of the dual honeycomb lattice ({\bf D}), and the degree distribution in this subgraph relates to quasiparticle behavior; $d=2$ sites correspond with closed loops or long chains of flipped spins, indicating collective spin flips or motion of fractional excitations.  {\bf E}, Field-dependence of the degree distribution indicates different field-induced kinetic regimes.}\label{fig:4}
\end{figure*}
\clearpage

\bibliography{kagome}

\begin{thebibliography}{47}
\expandafter\ifx\csname natexlab\endcsname\relax\def\natexlab#1{#1}\fi
\expandafter\ifx\csname bibnamefont\endcsname\relax
  \def\bibnamefont#1{#1}\fi
\expandafter\ifx\csname bibfnamefont\endcsname\relax
  \def\bibfnamefont#1{#1}\fi
\expandafter\ifx\csname citenamefont\endcsname\relax
  \def\citenamefont#1{#1}\fi
\expandafter\ifx\csname url\endcsname\relax
  \def\url#1{\texttt{#1}}\fi
\expandafter\ifx\csname urlprefix\endcsname\relax\def\urlprefix{URL }\fi
\providecommand{\bibinfo}[2]{#2}
\providecommand{\eprint}[2][]{\url{#2}}

\bibitem[{\citenamefont{Nelson}(2002)}]{nelson2002defects}
\bibinfo{author}{\bibfnamefont{D.~R.} \bibnamefont{Nelson}},
  \emph{\bibinfo{title}{Defects and geometry in condensed matter physics}}
  (\bibinfo{publisher}{Cambridge University Press}, \bibinfo{year}{2002}).

\bibitem[{\citenamefont{Henley}(2011)}]{henley2011classical}
\bibinfo{author}{\bibfnamefont{C.~L.} \bibnamefont{Henley}},
  \bibinfo{journal}{Journal of Physics: Condensed Matter}
  \textbf{\bibinfo{volume}{23}}, \bibinfo{pages}{164212}
  (\bibinfo{year}{2011}).

\bibitem[{\citenamefont{Matsuhira et~al.}(2002)\citenamefont{Matsuhira, Hiroi,
  Tayama, Takagi, and Sakakibara}}]{matsuhira2002new}
\bibinfo{author}{\bibfnamefont{K.}~\bibnamefont{Matsuhira}},
  \bibinfo{author}{\bibfnamefont{Z.}~\bibnamefont{Hiroi}},
  \bibinfo{author}{\bibfnamefont{T.}~\bibnamefont{Tayama}},
  \bibinfo{author}{\bibfnamefont{S.}~\bibnamefont{Takagi}}, \bibnamefont{and}
  \bibinfo{author}{\bibfnamefont{T.}~\bibnamefont{Sakakibara}},
  \bibinfo{journal}{Journal of Physics: Condensed Matter}
  \textbf{\bibinfo{volume}{14}}, \bibinfo{pages}{L559} (\bibinfo{year}{2002}),
  ISSN \bibinfo{issn}{0953-8984}.

\bibitem[{\citenamefont{Bramwell and Harris}(2020)}]{bramwell2020history}
\bibinfo{author}{\bibfnamefont{S.~T.} \bibnamefont{Bramwell}} \bibnamefont{and}
  \bibinfo{author}{\bibfnamefont{M.~J.} \bibnamefont{Harris}},
  \bibinfo{journal}{Journal of Physics Condensed Matter}
  \textbf{\bibinfo{volume}{32}} (\bibinfo{year}{2020}).

\bibitem[{\citenamefont{Perrin et~al.}(2016)\citenamefont{Perrin, Canals, and
  Rougemaille}}]{perrin2016extensive}
\bibinfo{author}{\bibfnamefont{Y.}~\bibnamefont{Perrin}},
  \bibinfo{author}{\bibfnamefont{B.}~\bibnamefont{Canals}}, \bibnamefont{and}
  \bibinfo{author}{\bibfnamefont{N.}~\bibnamefont{Rougemaille}},
  \bibinfo{journal}{Nature} \textbf{\bibinfo{volume}{540}},
  \bibinfo{pages}{410} (\bibinfo{year}{2016}).

\bibitem[{\citenamefont{Farhan et~al.}(2019)\citenamefont{Farhan, Saccone,
  Petersen, Dhuey, Chopdekar, Huang, Kent, Chen, Alava, Lippert
  et~al.}}]{farhan2019emergent}
\bibinfo{author}{\bibfnamefont{A.}~\bibnamefont{Farhan}},
  \bibinfo{author}{\bibfnamefont{M.}~\bibnamefont{Saccone}},
  \bibinfo{author}{\bibfnamefont{C.~F.} \bibnamefont{Petersen}},
  \bibinfo{author}{\bibfnamefont{S.}~\bibnamefont{Dhuey}},
  \bibinfo{author}{\bibfnamefont{R.~V.} \bibnamefont{Chopdekar}},
  \bibinfo{author}{\bibfnamefont{Y.~L.} \bibnamefont{Huang}},
  \bibinfo{author}{\bibfnamefont{N.}~\bibnamefont{Kent}},
  \bibinfo{author}{\bibfnamefont{Z.}~\bibnamefont{Chen}},
  \bibinfo{author}{\bibfnamefont{M.~J.} \bibnamefont{Alava}},
  \bibinfo{author}{\bibfnamefont{T.}~\bibnamefont{Lippert}},
  \bibnamefont{et~al.}, \bibinfo{journal}{Science Advances}
  \textbf{\bibinfo{volume}{5}}, \bibinfo{pages}{1} (\bibinfo{year}{2019}).

\bibitem[{\citenamefont{King et~al.}(2021{\natexlab{a}})\citenamefont{King,
  Nisoli, Dahl, Poulin-Lamarre, and Lopez-Bezanilla}}]{King2021}
\bibinfo{author}{\bibfnamefont{A.~D.} \bibnamefont{King}},
  \bibinfo{author}{\bibfnamefont{C.}~\bibnamefont{Nisoli}},
  \bibinfo{author}{\bibfnamefont{E.~D.} \bibnamefont{Dahl}},
  \bibinfo{author}{\bibfnamefont{G.}~\bibnamefont{Poulin-Lamarre}},
  \bibnamefont{and}
  \bibinfo{author}{\bibfnamefont{A.}~\bibnamefont{Lopez-Bezanilla}},
  \bibinfo{journal}{Science} \textbf{\bibinfo{volume}{373}},
  \bibinfo{pages}{576} (\bibinfo{year}{2021}{\natexlab{a}}).

\bibitem[{\citenamefont{Pauling}(1935)}]{pauling1935structure}
\bibinfo{author}{\bibfnamefont{L.}~\bibnamefont{Pauling}},
  \bibinfo{journal}{Journal of the American Chemical Society}
  \textbf{\bibinfo{volume}{57}}, \bibinfo{pages}{2680} (\bibinfo{year}{1935}),
  ISSN \bibinfo{issn}{0002-7863}.

\bibitem[{\citenamefont{Castelnovo et~al.}(2008)\citenamefont{Castelnovo,
  Moessner, and Sondhi}}]{Castelnovo2008}
\bibinfo{author}{\bibfnamefont{C.}~\bibnamefont{Castelnovo}},
  \bibinfo{author}{\bibfnamefont{R.}~\bibnamefont{Moessner}}, \bibnamefont{and}
  \bibinfo{author}{\bibfnamefont{S.~L.} \bibnamefont{Sondhi}},
  \bibinfo{journal}{Nature} \textbf{\bibinfo{volume}{451}}, \bibinfo{pages}{42}
  (\bibinfo{year}{2008}).

\bibitem[{\citenamefont{{Sriram Shastry} and
  Sutherland}(1981)}]{shastry1981exact}
\bibinfo{author}{\bibfnamefont{B.}~\bibnamefont{{Sriram Shastry}}}
  \bibnamefont{and}
  \bibinfo{author}{\bibfnamefont{B.}~\bibnamefont{Sutherland}},
  \bibinfo{journal}{Physica B+C} \textbf{\bibinfo{volume}{108}},
  \bibinfo{pages}{1069} (\bibinfo{year}{1981}), ISSN \bibinfo{issn}{03784363}.

\bibitem[{\citenamefont{Wills et~al.}(2002)\citenamefont{Wills, Ballou, and
  Lacroix}}]{wills2002model}
\bibinfo{author}{\bibfnamefont{A.}~\bibnamefont{Wills}},
  \bibinfo{author}{\bibfnamefont{R.}~\bibnamefont{Ballou}}, \bibnamefont{and}
  \bibinfo{author}{\bibfnamefont{C.}~\bibnamefont{Lacroix}},
  \bibinfo{journal}{Physical Review B} \textbf{\bibinfo{volume}{66}},
  \bibinfo{pages}{144407} (\bibinfo{year}{2002}).

\bibitem[{\citenamefont{Moessner and Sondhi}(2003)}]{moessner2003theory}
\bibinfo{author}{\bibfnamefont{R.}~\bibnamefont{Moessner}} \bibnamefont{and}
  \bibinfo{author}{\bibfnamefont{S.~L.} \bibnamefont{Sondhi}},
  \bibinfo{journal}{Physical Review B} \textbf{\bibinfo{volume}{68}},
  \bibinfo{pages}{064411} (\bibinfo{year}{2003}).

\bibitem[{\citenamefont{M\"{o}ller and Moessner}(2009)}]{Moller2009}
\bibinfo{author}{\bibfnamefont{G.}~\bibnamefont{M\"{o}ller}} \bibnamefont{and}
  \bibinfo{author}{\bibfnamefont{R.}~\bibnamefont{Moessner}},
  \bibinfo{journal}{Phys. Rev. B} \textbf{\bibinfo{volume}{80}},
  \bibinfo{pages}{140409} (\bibinfo{year}{2009}).

\bibitem[{\citenamefont{Chern et~al.}(2011)\citenamefont{Chern, Mellado, and
  Tchernyshyov}}]{Chern2011}
\bibinfo{author}{\bibfnamefont{G.-W.} \bibnamefont{Chern}},
  \bibinfo{author}{\bibfnamefont{P.}~\bibnamefont{Mellado}}, \bibnamefont{and}
  \bibinfo{author}{\bibfnamefont{O.}~\bibnamefont{Tchernyshyov}},
  \bibinfo{journal}{Phys. Rev. Lett.} \textbf{\bibinfo{volume}{106}},
  \bibinfo{pages}{207202} (\bibinfo{year}{2011}).

\bibitem[{\citenamefont{Lib\'al et~al.}(2018)\citenamefont{Lib\'al, Nisoli,
  Reichhardt, and Reichhardt}}]{libal2017}
\bibinfo{author}{\bibfnamefont{A.}~\bibnamefont{Lib\'al}},
  \bibinfo{author}{\bibfnamefont{C.}~\bibnamefont{Nisoli}},
  \bibinfo{author}{\bibfnamefont{C.~J.~O.} \bibnamefont{Reichhardt}},
  \bibnamefont{and}
  \bibinfo{author}{\bibfnamefont{C.}~\bibnamefont{Reichhardt}},
  \bibinfo{journal}{Phys. Rev. Lett.} \textbf{\bibinfo{volume}{120}},
  \bibinfo{pages}{027204} (\bibinfo{year}{2018}),
  \urlprefix\url{https://link.aps.org/doi/10.1103/PhysRevLett.120.027204}.

\bibitem[{\citenamefont{Balents}(2010)}]{balents2010spin}
\bibinfo{author}{\bibfnamefont{L.}~\bibnamefont{Balents}},
  \bibinfo{journal}{Nature} \textbf{\bibinfo{volume}{464}},
  \bibinfo{pages}{199} (\bibinfo{year}{2010}).

\bibitem[{\citenamefont{Raban et~al.}(2019)\citenamefont{Raban, Suen, Berthier,
  and Holdsworth}}]{raban2019multiple}
\bibinfo{author}{\bibfnamefont{V.}~\bibnamefont{Raban}},
  \bibinfo{author}{\bibfnamefont{C.}~\bibnamefont{Suen}},
  \bibinfo{author}{\bibfnamefont{L.}~\bibnamefont{Berthier}}, \bibnamefont{and}
  \bibinfo{author}{\bibfnamefont{P.}~\bibnamefont{Holdsworth}},
  \bibinfo{journal}{Physical Review B} \textbf{\bibinfo{volume}{99}},
  \bibinfo{pages}{224425} (\bibinfo{year}{2019}).

\bibitem[{\citenamefont{Lhotel et~al.}(2020)\citenamefont{Lhotel, Jaubert, and
  Holdsworth}}]{lhotel2020fragmentation}
\bibinfo{author}{\bibfnamefont{E.}~\bibnamefont{Lhotel}},
  \bibinfo{author}{\bibfnamefont{L.~D.} \bibnamefont{Jaubert}},
  \bibnamefont{and} \bibinfo{author}{\bibfnamefont{P.~C.}
  \bibnamefont{Holdsworth}}, \bibinfo{journal}{Journal of Low Temperature
  Physics} \textbf{\bibinfo{volume}{201}}, \bibinfo{pages}{710}
  (\bibinfo{year}{2020}).

\bibitem[{\citenamefont{Qi et~al.}(2008)\citenamefont{Qi, Brintlinger, and
  Cumings}}]{qi2008direct}
\bibinfo{author}{\bibfnamefont{Y.}~\bibnamefont{Qi}},
  \bibinfo{author}{\bibfnamefont{T.}~\bibnamefont{Brintlinger}},
  \bibnamefont{and} \bibinfo{author}{\bibfnamefont{J.}~\bibnamefont{Cumings}},
  \bibinfo{journal}{Physical Review B} \textbf{\bibinfo{volume}{77}},
  \bibinfo{pages}{1} (\bibinfo{year}{2008}).

\bibitem[{\citenamefont{Lib\'al et~al.}(2009)\citenamefont{Lib\'al, Reichhardt,
  and Reichhardt}}]{libal2009}
\bibinfo{author}{\bibfnamefont{A.}~\bibnamefont{Lib\'al}},
  \bibinfo{author}{\bibfnamefont{C.~J.~O.} \bibnamefont{Reichhardt}},
  \bibnamefont{and}
  \bibinfo{author}{\bibfnamefont{C.}~\bibnamefont{Reichhardt}},
  \bibinfo{journal}{Phys. Rev. Lett.} \textbf{\bibinfo{volume}{102}},
  \bibinfo{pages}{237004} (\bibinfo{year}{2009}).

\bibitem[{\citenamefont{Wang et~al.}(2018)\citenamefont{Wang, Ma, Xu, Xiao,
  Snezhko, Divan, Ocola, Pearson, Janko, and Kwok}}]{wang2018switchable}
\bibinfo{author}{\bibfnamefont{Y.-L.} \bibnamefont{Wang}},
  \bibinfo{author}{\bibfnamefont{X.}~\bibnamefont{Ma}},
  \bibinfo{author}{\bibfnamefont{J.}~\bibnamefont{Xu}},
  \bibinfo{author}{\bibfnamefont{Z.-L.} \bibnamefont{Xiao}},
  \bibinfo{author}{\bibfnamefont{A.}~\bibnamefont{Snezhko}},
  \bibinfo{author}{\bibfnamefont{R.}~\bibnamefont{Divan}},
  \bibinfo{author}{\bibfnamefont{L.~E.} \bibnamefont{Ocola}},
  \bibinfo{author}{\bibfnamefont{J.~E.} \bibnamefont{Pearson}},
  \bibinfo{author}{\bibfnamefont{B.}~\bibnamefont{Janko}}, \bibnamefont{and}
  \bibinfo{author}{\bibfnamefont{W.-K.} \bibnamefont{Kwok}},
  \bibinfo{journal}{Nature Nanotechnology} \textbf{\bibinfo{volume}{13}},
  \bibinfo{pages}{560} (\bibinfo{year}{2018}), ISSN \bibinfo{issn}{1748-3387}.

\bibitem[{\citenamefont{Xue et~al.}(2018)\citenamefont{Xue, Ge, He, Zharinov,
  Moshchalkov, Zhou, Silhanek, and Van~de Vondel}}]{xue2018tunable}
\bibinfo{author}{\bibfnamefont{C.}~\bibnamefont{Xue}},
  \bibinfo{author}{\bibfnamefont{J.-Y.} \bibnamefont{Ge}},
  \bibinfo{author}{\bibfnamefont{A.}~\bibnamefont{He}},
  \bibinfo{author}{\bibfnamefont{V.~S.} \bibnamefont{Zharinov}},
  \bibinfo{author}{\bibfnamefont{V.~V.} \bibnamefont{Moshchalkov}},
  \bibinfo{author}{\bibfnamefont{Y.~H.} \bibnamefont{Zhou}},
  \bibinfo{author}{\bibfnamefont{A.~V.} \bibnamefont{Silhanek}},
  \bibnamefont{and} \bibinfo{author}{\bibfnamefont{J.}~\bibnamefont{Van~de
  Vondel}}, \bibinfo{journal}{Phys. Rev. B} \textbf{\bibinfo{volume}{97}},
  \bibinfo{pages}{134506} (\bibinfo{year}{2018}).

\bibitem[{\citenamefont{Mellado et~al.}(2012)\citenamefont{Mellado, Concha, and
  Mahadevan}}]{Mellado2012}
\bibinfo{author}{\bibfnamefont{P.}~\bibnamefont{Mellado}},
  \bibinfo{author}{\bibfnamefont{A.}~\bibnamefont{Concha}}, \bibnamefont{and}
  \bibinfo{author}{\bibfnamefont{L.}~\bibnamefont{Mahadevan}},
  \bibinfo{journal}{Physical Review Letters} \textbf{\bibinfo{volume}{109}},
  \bibinfo{pages}{257203} (\bibinfo{year}{2012}).

\bibitem[{\citenamefont{Duzgun and Nisoli}(2021)}]{duzgun2021skyrmion}
\bibinfo{author}{\bibfnamefont{A.}~\bibnamefont{Duzgun}} \bibnamefont{and}
  \bibinfo{author}{\bibfnamefont{C.}~\bibnamefont{Nisoli}},
  \bibinfo{journal}{Phys. Rev. Lett.} \textbf{\bibinfo{volume}{126}},
  \bibinfo{pages}{047801} (\bibinfo{year}{2021}).

\bibitem[{\citenamefont{Zhao et~al.}(2020)\citenamefont{Zhao, Deng, Chen, Ross,
  Pet{\v{r}}{\'\i}{\v{c}}ek, G{\"u}nther, Russina, Hutanu, and
  Gegenwart}}]{zhao2020realization}
\bibinfo{author}{\bibfnamefont{K.}~\bibnamefont{Zhao}},
  \bibinfo{author}{\bibfnamefont{H.}~\bibnamefont{Deng}},
  \bibinfo{author}{\bibfnamefont{H.}~\bibnamefont{Chen}},
  \bibinfo{author}{\bibfnamefont{K.~A.} \bibnamefont{Ross}},
  \bibinfo{author}{\bibfnamefont{V.}~\bibnamefont{Pet{\v{r}}{\'\i}{\v{c}}ek}},
  \bibinfo{author}{\bibfnamefont{G.}~\bibnamefont{G{\"u}nther}},
  \bibinfo{author}{\bibfnamefont{M.}~\bibnamefont{Russina}},
  \bibinfo{author}{\bibfnamefont{V.}~\bibnamefont{Hutanu}}, \bibnamefont{and}
  \bibinfo{author}{\bibfnamefont{P.}~\bibnamefont{Gegenwart}},
  \bibinfo{journal}{Science} \textbf{\bibinfo{volume}{367}},
  \bibinfo{pages}{1218} (\bibinfo{year}{2020}).

\bibitem[{\citenamefont{Hua et~al.}(2021)\citenamefont{Hua, Xia, Wang, Li, Liu,
  Wu, Wang, Li, Ding, Hu et~al.}}]{hua2021highly}
\bibinfo{author}{\bibfnamefont{M.}~\bibnamefont{Hua}},
  \bibinfo{author}{\bibfnamefont{B.}~\bibnamefont{Xia}},
  \bibinfo{author}{\bibfnamefont{M.}~\bibnamefont{Wang}},
  \bibinfo{author}{\bibfnamefont{E.}~\bibnamefont{Li}},
  \bibinfo{author}{\bibfnamefont{J.}~\bibnamefont{Liu}},
  \bibinfo{author}{\bibfnamefont{T.}~\bibnamefont{Wu}},
  \bibinfo{author}{\bibfnamefont{Y.}~\bibnamefont{Wang}},
  \bibinfo{author}{\bibfnamefont{R.}~\bibnamefont{Li}},
  \bibinfo{author}{\bibfnamefont{H.}~\bibnamefont{Ding}},
  \bibinfo{author}{\bibfnamefont{J.}~\bibnamefont{Hu}}, \bibnamefont{et~al.},
  \bibinfo{journal}{The Journal of Physical Chemistry Letters}
  \textbf{\bibinfo{volume}{12}}, \bibinfo{pages}{3733} (\bibinfo{year}{2021}),
  ISSN \bibinfo{issn}{1948-7185}.

\bibitem[{\citenamefont{Meeussen et~al.}(2020)\citenamefont{Meeussen, Oğuz,
  Shokef, and van Hecke}}]{meeussen2020topological}
\bibinfo{author}{\bibfnamefont{A.~S.} \bibnamefont{Meeussen}},
  \bibinfo{author}{\bibfnamefont{E.~C.} \bibnamefont{Oğuz}},
  \bibinfo{author}{\bibfnamefont{Y.}~\bibnamefont{Shokef}}, \bibnamefont{and}
  \bibinfo{author}{\bibfnamefont{M.}~\bibnamefont{van Hecke}},
  \bibinfo{journal}{Nature Physics} \textbf{\bibinfo{volume}{16}},
  \bibinfo{pages}{307} (\bibinfo{year}{2020}), ISSN \bibinfo{issn}{1745-2473}.

\bibitem[{\citenamefont{Pisanty et~al.}(2021)\citenamefont{Pisanty, Oguz,
  Nisoli, and Shokef}}]{pisanty2020topological}
\bibinfo{author}{\bibfnamefont{B.}~\bibnamefont{Pisanty}},
  \bibinfo{author}{\bibfnamefont{E.~C.} \bibnamefont{Oguz}},
  \bibinfo{author}{\bibfnamefont{C.}~\bibnamefont{Nisoli}}, \bibnamefont{and}
  \bibinfo{author}{\bibfnamefont{Y.}~\bibnamefont{Shokef}},
  \bibinfo{journal}{SciPost Phys.} \textbf{\bibinfo{volume}{10}},
  \bibinfo{pages}{136} (\bibinfo{year}{2021}).

\bibitem[{\citenamefont{Macdonald et~al.}(2011)\citenamefont{Macdonald,
  Holdsworth, and Melko}}]{macdonald2011classical}
\bibinfo{author}{\bibfnamefont{A.~J.} \bibnamefont{Macdonald}},
  \bibinfo{author}{\bibfnamefont{P.~C.~W.} \bibnamefont{Holdsworth}},
  \bibnamefont{and} \bibinfo{author}{\bibfnamefont{R.~G.} \bibnamefont{Melko}},
  \bibinfo{journal}{Journal of Physics: Condensed Matter}
  \textbf{\bibinfo{volume}{23}}, \bibinfo{pages}{164208}
  (\bibinfo{year}{2011}), ISSN \bibinfo{issn}{0953-8984}.

\bibitem[{\citenamefont{Zhang et~al.}(2013)\citenamefont{Zhang, Gilbert,
  Nisoli, Chern, Erickson, O'Brien, Leighton, Lammert, Crespi, and
  Schiffer}}]{Zhang2013}
\bibinfo{author}{\bibfnamefont{S.}~\bibnamefont{Zhang}},
  \bibinfo{author}{\bibfnamefont{I.}~\bibnamefont{Gilbert}},
  \bibinfo{author}{\bibfnamefont{C.}~\bibnamefont{Nisoli}},
  \bibinfo{author}{\bibfnamefont{G.~W.} \bibnamefont{Chern}},
  \bibinfo{author}{\bibfnamefont{M.~J.} \bibnamefont{Erickson}},
  \bibinfo{author}{\bibfnamefont{L.}~\bibnamefont{O'Brien}},
  \bibinfo{author}{\bibfnamefont{C.}~\bibnamefont{Leighton}},
  \bibinfo{author}{\bibfnamefont{P.~E.} \bibnamefont{Lammert}},
  \bibinfo{author}{\bibfnamefont{V.~H.} \bibnamefont{Crespi}},
  \bibnamefont{and} \bibinfo{author}{\bibfnamefont{P.}~\bibnamefont{Schiffer}},
  \bibinfo{journal}{Nature} \textbf{\bibinfo{volume}{500}},
  \bibinfo{pages}{553} (\bibinfo{year}{2013}), ISSN \bibinfo{issn}{00280836}.

\bibitem[{\citenamefont{Drisko et~al.}(2015)\citenamefont{Drisko, Daunheimer,
  and Cumings}}]{drisko2015fepd}
\bibinfo{author}{\bibfnamefont{J.}~\bibnamefont{Drisko}},
  \bibinfo{author}{\bibfnamefont{S.}~\bibnamefont{Daunheimer}},
  \bibnamefont{and} \bibinfo{author}{\bibfnamefont{J.}~\bibnamefont{Cumings}},
  \bibinfo{journal}{Physical Review B} \textbf{\bibinfo{volume}{91}},
  \bibinfo{pages}{224406} (\bibinfo{year}{2015}).

\bibitem[{\citenamefont{Levis et~al.}(2013)\citenamefont{Levis, Cugliandolo,
  Foini, and Tarzia}}]{levis2013thermal}
\bibinfo{author}{\bibfnamefont{D.}~\bibnamefont{Levis}},
  \bibinfo{author}{\bibfnamefont{L.~F.} \bibnamefont{Cugliandolo}},
  \bibinfo{author}{\bibfnamefont{L.}~\bibnamefont{Foini}}, \bibnamefont{and}
  \bibinfo{author}{\bibfnamefont{M.}~\bibnamefont{Tarzia}},
  \bibinfo{journal}{Physical review letters} \textbf{\bibinfo{volume}{110}},
  \bibinfo{pages}{207206} (\bibinfo{year}{2013}).

\bibitem[{\citenamefont{Anghinolfi et~al.}(2015)\citenamefont{Anghinolfi,
  Luetkens, Perron, Flokstra, Sendetskyi, Suter, Prokscha, Derlet, Lee, and
  Heyderman}}]{anghinolfi2015thermodynamic}
\bibinfo{author}{\bibfnamefont{L.}~\bibnamefont{Anghinolfi}},
  \bibinfo{author}{\bibfnamefont{H.}~\bibnamefont{Luetkens}},
  \bibinfo{author}{\bibfnamefont{J.}~\bibnamefont{Perron}},
  \bibinfo{author}{\bibfnamefont{M.}~\bibnamefont{Flokstra}},
  \bibinfo{author}{\bibfnamefont{O.}~\bibnamefont{Sendetskyi}},
  \bibinfo{author}{\bibfnamefont{A.}~\bibnamefont{Suter}},
  \bibinfo{author}{\bibfnamefont{T.}~\bibnamefont{Prokscha}},
  \bibinfo{author}{\bibfnamefont{P.}~\bibnamefont{Derlet}},
  \bibinfo{author}{\bibfnamefont{S.}~\bibnamefont{Lee}}, \bibnamefont{and}
  \bibinfo{author}{\bibfnamefont{L.}~\bibnamefont{Heyderman}},
  \bibinfo{journal}{Nature communications} \textbf{\bibinfo{volume}{6}}
  (\bibinfo{year}{2015}).

\bibitem[{\citenamefont{Zhang et~al.}(2012)\citenamefont{Zhang, Li, Gilbert,
  Bartell, Erickson, Pan, Lammert, Nisoli, Kohli, Misra
  et~al.}}]{zhang2012perpendicular}
\bibinfo{author}{\bibfnamefont{S.}~\bibnamefont{Zhang}},
  \bibinfo{author}{\bibfnamefont{J.}~\bibnamefont{Li}},
  \bibinfo{author}{\bibfnamefont{I.}~\bibnamefont{Gilbert}},
  \bibinfo{author}{\bibfnamefont{J.}~\bibnamefont{Bartell}},
  \bibinfo{author}{\bibfnamefont{M.~J.} \bibnamefont{Erickson}},
  \bibinfo{author}{\bibfnamefont{Y.}~\bibnamefont{Pan}},
  \bibinfo{author}{\bibfnamefont{P.~E.} \bibnamefont{Lammert}},
  \bibinfo{author}{\bibfnamefont{C.}~\bibnamefont{Nisoli}},
  \bibinfo{author}{\bibfnamefont{K.}~\bibnamefont{Kohli}},
  \bibinfo{author}{\bibfnamefont{R.}~\bibnamefont{Misra}},
  \bibnamefont{et~al.}, \bibinfo{journal}{Physical review letters}
  \textbf{\bibinfo{volume}{109}}, \bibinfo{pages}{087201}
  (\bibinfo{year}{2012}).

\bibitem[{\citenamefont{Chamon et~al.}(2020)\citenamefont{Chamon, Green, and
  Yang}}]{Chamon2020}
\bibinfo{author}{\bibfnamefont{C.}~\bibnamefont{Chamon}},
  \bibinfo{author}{\bibfnamefont{D.}~\bibnamefont{Green}}, \bibnamefont{and}
  \bibinfo{author}{\bibfnamefont{Z.-C.} \bibnamefont{Yang}},
  \bibinfo{journal}{Physical Review Letters} \textbf{\bibinfo{volume}{125}},
  \bibinfo{pages}{067203} (\bibinfo{year}{2020}), ISSN
  \bibinfo{issn}{0031-9007}.

\bibitem[{\citenamefont{King et~al.}(2021{\natexlab{b}})\citenamefont{King,
  Batista, Raymond, Lanting, Ozfidan, Poulin-Lamarre, Zhang, and
  Amin}}]{King2021prxq}
\bibinfo{author}{\bibfnamefont{A.~D.} \bibnamefont{King}},
  \bibinfo{author}{\bibfnamefont{C.~D.} \bibnamefont{Batista}},
  \bibinfo{author}{\bibfnamefont{J.}~\bibnamefont{Raymond}},
  \bibinfo{author}{\bibfnamefont{T.}~\bibnamefont{Lanting}},
  \bibinfo{author}{\bibfnamefont{I.}~\bibnamefont{Ozfidan}},
  \bibinfo{author}{\bibfnamefont{G.}~\bibnamefont{Poulin-Lamarre}},
  \bibinfo{author}{\bibfnamefont{H.}~\bibnamefont{Zhang}}, \bibnamefont{and}
  \bibinfo{author}{\bibfnamefont{M.~H.} \bibnamefont{Amin}},
  \bibinfo{journal}{PRX Quantum} \textbf{\bibinfo{volume}{2}},
  \bibinfo{pages}{030317} (\bibinfo{year}{2021}{\natexlab{b}}).

\bibitem[{\citenamefont{Rougemaille et~al.}(2011)\citenamefont{Rougemaille,
  Montaigne, Canals, Duluard, Lacour, Hehn, Belkhou, Fruchart, El~Moussaoui,
  Bendounan et~al.}}]{Rougemaille2011}
\bibinfo{author}{\bibfnamefont{N.}~\bibnamefont{Rougemaille}},
  \bibinfo{author}{\bibfnamefont{F.}~\bibnamefont{Montaigne}},
  \bibinfo{author}{\bibfnamefont{B.}~\bibnamefont{Canals}},
  \bibinfo{author}{\bibfnamefont{A.}~\bibnamefont{Duluard}},
  \bibinfo{author}{\bibfnamefont{D.}~\bibnamefont{Lacour}},
  \bibinfo{author}{\bibfnamefont{M.}~\bibnamefont{Hehn}},
  \bibinfo{author}{\bibfnamefont{R.}~\bibnamefont{Belkhou}},
  \bibinfo{author}{\bibfnamefont{O.}~\bibnamefont{Fruchart}},
  \bibinfo{author}{\bibfnamefont{S.}~\bibnamefont{El~Moussaoui}},
  \bibinfo{author}{\bibfnamefont{A.}~\bibnamefont{Bendounan}},
  \bibnamefont{et~al.}, \bibinfo{journal}{Phys. Rev. Lett.}
  \textbf{\bibinfo{volume}{106}}, \bibinfo{pages}{057209}
  (\bibinfo{year}{2011}).

\bibitem[{\citenamefont{Lammert et~al.}(2010)\citenamefont{Lammert, Ke, Li,
  Nisoli, Garand, Crespi, and Schiffer}}]{Lammert2010}
\bibinfo{author}{\bibfnamefont{P.~E.} \bibnamefont{Lammert}},
  \bibinfo{author}{\bibfnamefont{X.}~\bibnamefont{Ke}},
  \bibinfo{author}{\bibfnamefont{J.}~\bibnamefont{Li}},
  \bibinfo{author}{\bibfnamefont{C.}~\bibnamefont{Nisoli}},
  \bibinfo{author}{\bibfnamefont{D.~M.} \bibnamefont{Garand}},
  \bibinfo{author}{\bibfnamefont{V.~H.} \bibnamefont{Crespi}},
  \bibnamefont{and} \bibinfo{author}{\bibfnamefont{P.}~\bibnamefont{Schiffer}},
  \bibinfo{journal}{Nat. Phys.} \textbf{\bibinfo{volume}{6}},
  \bibinfo{pages}{786} (\bibinfo{year}{2010}).

\bibitem[{\citenamefont{Lamberty et~al.}(2013)\citenamefont{Lamberty,
  Papanikolaou, and Henley}}]{lamberty2013classical}
\bibinfo{author}{\bibfnamefont{R.~Z.} \bibnamefont{Lamberty}},
  \bibinfo{author}{\bibfnamefont{S.}~\bibnamefont{Papanikolaou}},
  \bibnamefont{and} \bibinfo{author}{\bibfnamefont{C.~L.}
  \bibnamefont{Henley}}, \bibinfo{journal}{Physical review letters}
  \textbf{\bibinfo{volume}{111}}, \bibinfo{pages}{245701}
  (\bibinfo{year}{2013}).

\bibitem[{\citenamefont{Lao et~al.}(2018)\citenamefont{Lao, Caravelli, Sheikh,
  Sklenar, Gardeazabal, Watts, Albrecht, Scholl, Dahmen, Nisoli
  et~al.}}]{lao2018classical}
\bibinfo{author}{\bibfnamefont{Y.}~\bibnamefont{Lao}},
  \bibinfo{author}{\bibfnamefont{F.}~\bibnamefont{Caravelli}},
  \bibinfo{author}{\bibfnamefont{M.}~\bibnamefont{Sheikh}},
  \bibinfo{author}{\bibfnamefont{J.}~\bibnamefont{Sklenar}},
  \bibinfo{author}{\bibfnamefont{D.}~\bibnamefont{Gardeazabal}},
  \bibinfo{author}{\bibfnamefont{J.~D.} \bibnamefont{Watts}},
  \bibinfo{author}{\bibfnamefont{A.~M.} \bibnamefont{Albrecht}},
  \bibinfo{author}{\bibfnamefont{A.}~\bibnamefont{Scholl}},
  \bibinfo{author}{\bibfnamefont{K.}~\bibnamefont{Dahmen}},
  \bibinfo{author}{\bibfnamefont{C.}~\bibnamefont{Nisoli}},
  \bibnamefont{et~al.}, \bibinfo{journal}{Nature Physics}
  \textbf{\bibinfo{volume}{14}}, \bibinfo{pages}{723} (\bibinfo{year}{2018}).

\bibitem[{\citenamefont{Canals et~al.}(2016)\citenamefont{Canals, Chioar,
  Nguyen, Hehn, Lacour, Montaigne, Locatelli, Mente{\c s}, Burgos, and
  Rougemaille}}]{canals2016fragmentation}
\bibinfo{author}{\bibfnamefont{B.}~\bibnamefont{Canals}},
  \bibinfo{author}{\bibfnamefont{I.-A.} \bibnamefont{Chioar}},
  \bibinfo{author}{\bibfnamefont{V.-D.} \bibnamefont{Nguyen}},
  \bibinfo{author}{\bibfnamefont{M.}~\bibnamefont{Hehn}},
  \bibinfo{author}{\bibfnamefont{D.}~\bibnamefont{Lacour}},
  \bibinfo{author}{\bibfnamefont{F.}~\bibnamefont{Montaigne}},
  \bibinfo{author}{\bibfnamefont{A.}~\bibnamefont{Locatelli}},
  \bibinfo{author}{\bibfnamefont{T.~O.} \bibnamefont{Mente{\c s}}},
  \bibinfo{author}{\bibfnamefont{B.~S.} \bibnamefont{Burgos}},
  \bibnamefont{and}
  \bibinfo{author}{\bibfnamefont{N.}~\bibnamefont{Rougemaille}},
  \bibinfo{journal}{Nature Communications} \textbf{\bibinfo{volume}{7}},
  \bibinfo{pages}{11446} (\bibinfo{year}{2016}), ISSN
  \bibinfo{issn}{2041-1723}.

\bibitem[{\citenamefont{Brooks-Bartlett
  et~al.}(2014)\citenamefont{Brooks-Bartlett, Banks, Jaubert, Harman-Clarke,
  and Holdsworth}}]{brooks2014magnetic}
\bibinfo{author}{\bibfnamefont{M.}~\bibnamefont{Brooks-Bartlett}},
  \bibinfo{author}{\bibfnamefont{S.~T.} \bibnamefont{Banks}},
  \bibinfo{author}{\bibfnamefont{L.~D.} \bibnamefont{Jaubert}},
  \bibinfo{author}{\bibfnamefont{A.}~\bibnamefont{Harman-Clarke}},
  \bibnamefont{and} \bibinfo{author}{\bibfnamefont{P.~C.}
  \bibnamefont{Holdsworth}}, \bibinfo{journal}{Physical Review X}
  \textbf{\bibinfo{volume}{4}}, \bibinfo{pages}{011007} (\bibinfo{year}{2014}).

\bibitem[{\citenamefont{Castelnovo and
  Chamon}(2007)}]{castelnovo2007topological}
\bibinfo{author}{\bibfnamefont{C.}~\bibnamefont{Castelnovo}} \bibnamefont{and}
  \bibinfo{author}{\bibfnamefont{C.}~\bibnamefont{Chamon}},
  \bibinfo{journal}{Physical Review B} \textbf{\bibinfo{volume}{76}},
  \bibinfo{pages}{174416} (\bibinfo{year}{2007}).

\bibitem[{\citenamefont{Moessner et~al.}(2000)\citenamefont{Moessner, Sondhi,
  and Chandra}}]{Moessner2000}
\bibinfo{author}{\bibfnamefont{R.}~\bibnamefont{Moessner}},
  \bibinfo{author}{\bibfnamefont{S.~L.} \bibnamefont{Sondhi}},
  \bibnamefont{and} \bibinfo{author}{\bibfnamefont{P.}~\bibnamefont{Chandra}},
  \bibinfo{journal}{Physical Review Letters} \textbf{\bibinfo{volume}{84}},
  \bibinfo{pages}{4457} (\bibinfo{year}{2000}), ISSN \bibinfo{issn}{0031-9007}.

\bibitem[{\citenamefont{Boothby et~al.}(2021)\citenamefont{Boothby, Enderud,
  Lanting, Molavi, Tsai, Volkmann, Altomare, Amin, Babcock, Berkley
  et~al.}}]{Boothby2021}
\bibinfo{author}{\bibfnamefont{K.}~\bibnamefont{Boothby}},
  \bibinfo{author}{\bibfnamefont{C.}~\bibnamefont{Enderud}},
  \bibinfo{author}{\bibfnamefont{T.}~\bibnamefont{Lanting}},
  \bibinfo{author}{\bibfnamefont{R.}~\bibnamefont{Molavi}},
  \bibinfo{author}{\bibfnamefont{N.}~\bibnamefont{Tsai}},
  \bibinfo{author}{\bibfnamefont{M.~H.} \bibnamefont{Volkmann}},
  \bibinfo{author}{\bibfnamefont{F.}~\bibnamefont{Altomare}},
  \bibinfo{author}{\bibfnamefont{M.~H.} \bibnamefont{Amin}},
  \bibinfo{author}{\bibfnamefont{M.}~\bibnamefont{Babcock}},
  \bibinfo{author}{\bibfnamefont{A.~J.} \bibnamefont{Berkley}},
  \bibnamefont{et~al.}, \emph{\bibinfo{title}{{Architectural considerations in
  the design of a third-generation superconducting quantum annealing
  processor}}} (\bibinfo{year}{2021}), \eprint{2108.02322}.

\bibitem[{\citenamefont{King et~al.}(2018)\citenamefont{King, Carrasquilla,
  Raymond, Ozfidan, Andriyash, Berkley, Reis, Lanting, Harris, Altomare
  et~al.}}]{King2018}
\bibinfo{author}{\bibfnamefont{A.~D.} \bibnamefont{King}},
  \bibinfo{author}{\bibfnamefont{J.}~\bibnamefont{Carrasquilla}},
  \bibinfo{author}{\bibfnamefont{J.}~\bibnamefont{Raymond}},
  \bibinfo{author}{\bibfnamefont{I.}~\bibnamefont{Ozfidan}},
  \bibinfo{author}{\bibfnamefont{E.}~\bibnamefont{Andriyash}},
  \bibinfo{author}{\bibfnamefont{A.~J.} \bibnamefont{Berkley}},
  \bibinfo{author}{\bibfnamefont{M.}~\bibnamefont{Reis}},
  \bibinfo{author}{\bibfnamefont{T.}~\bibnamefont{Lanting}},
  \bibinfo{author}{\bibfnamefont{R.}~\bibnamefont{Harris}},
  \bibinfo{author}{\bibfnamefont{F.}~\bibnamefont{Altomare}},
  \bibnamefont{et~al.}, \bibinfo{journal}{Nature}
  \textbf{\bibinfo{volume}{560}}, \bibinfo{pages}{456} (\bibinfo{year}{2018}).

\bibitem[{\citenamefont{Boothby et~al.}(2020)\citenamefont{Boothby, Bunyk,
  Raymond, and Roy}}]{Boothby2020}
\bibinfo{author}{\bibfnamefont{K.}~\bibnamefont{Boothby}},
  \bibinfo{author}{\bibfnamefont{P.}~\bibnamefont{Bunyk}},
  \bibinfo{author}{\bibfnamefont{J.}~\bibnamefont{Raymond}}, \bibnamefont{and}
  \bibinfo{author}{\bibfnamefont{A.}~\bibnamefont{Roy}},
  \emph{\bibinfo{title}{{Next-Generation Topology of D-Wave Quantum
  Processors}}} (\bibinfo{year}{2020}), \eprint{2003.00133},
  \urlprefix\url{http://arxiv.org/abs/2003.00133}.

\end{thebibliography}

\section*{Acknowledgments}

The authors aknowledge the contributions of technical staff at D-Wave, without whom this work would not be possible.  The work of ALB and CN was carried out under the auspices of the U.S. DoE through the Los Alamos National Laboratory, operated by Triad National Security, LLC (Contract No.~892333218NCA000001). JC acknowledges support from the Natural Sciences and Engineering Research Council (NSERC), the Shared Hierarchical Academic Research Computing Network (SHARCNET), Compute Canada, Google Quantum Research Award, and the Canadian Institute for Advanced Research (CIFAR) AI chair program. Resources used in preparing this research were provided, in part, by the Province of Ontario, the Government of Canada through CIFAR, and companies sponsoring the Vector Institute.

\section*{Competing interests}
The authors declare no competing interests.

\section*{Author Contributions}

J.C. first proposed the idea of a Kagome embedding in a D-Wave QA to A.D.K.. 
A.D.K., A.L.-B., C.N., and J.C. conceived the project. 
C.N., A.D.K. contributed to the design of the experiments. 
A.D.K. and K.B. realized the embedding.  
J.R. performed supporting measurements. 
A.D.K., A.L.-B. performed QA experiments. 
A.D.K. performed data analysis. 
C.N. provided the theoretical framework for experiment design and result interpretation.
C.N. drafted the manuscript with A.D.K. 
All authors contributed to the final version of the manuscript.

\section*{Methods}


The QA processor used in this work was a D-Wave Advantage QPU (Fig.~1d) housed in Burnaby, BC, Canada, operating at $T=\SI{12}{mK}$ and accessed remotely.  The QPU contains 5627 operable superconducting flux qubits of which we used 2739 to implement our kagome qubit spin ice.  The architecture is discussed in Ref.~\cite{Boothby2021}.

In quantum annealing, the Hamiltonian (1) in the Main Text is controlled by an annealing parameter $s$ ranging from $0$ to $1$:
\begin{equation}\label{eq:tfim_s}
  H_Q(s) = -\Gamma(s)\sum_{i}\sigma_i^x + \mathcal J(s)\big(\sum_ih_i\sigma_i^z +\sum_{ij}J_{ij}\sigma_i^z\sigma_j^z \big),
\end{equation}
where $\Gamma(0)\gg \mathcal J(0)$ and $\Gamma(1)\approx 0 \ll \mathcal J(1)$.

Thus a typical ``forward anneal'', in which $s$ is ramped linearly for the duration of anneal time $t_a$ ($s=t/t_a$) begins in an easily-prepared superposition ground state and ends in a low-energy state of a classical Ising Hamiltonian. For simulating spin systems, it has proven useful \cite{King2018, King2021} to employ a ``quantum evolution Monte Carlo'' method, in which a chain of classical samples $S_0,\ldots S_k$ is generated.  To generate $S_i$, the system is initialized in state $S_0$ at the end of the anneal ($s=1$), then ``reverse annealed'' back to some intermediate $s^*$, paused at $s^*$ to allow equilibration for some time $t_p$, then quickly quenched back to $s=1$.  Although this method can be used to estimate observables from a transverse field Ising model at $s^*$ \cite{King2018}, here we just use quantum fluctuations as a driver of mixing dynamics between low energy state in the kagome ice system.

In this work we generate chains of $k=128$ samples, starting with a random initial state $S_0$.  To estimate equilibrium properties (Fig.~3) we use $t_p=\SI{256}{\micro s}$ and discard the first $64$ samples of each chain (and the random initial state) as Monte Carlo burn-in.  For dynamics inquiries (Fig.~4) we use $t_p=\SI{1}{\micro s}$.  In both cases we interrogate Hamiltonian (\ref{eq:tfim_s}) using $s^*=0.32$, which was chosen to give an appropriate amount of mixing in one microsecond (smaller $s$ leads to faster mixing since both $\Gamma/\mathcal J$ and $T/\mathcal J$ are larger \cite{King2021}).  When statistical quantities are estimated, we take the average of $200$ repeated experimental iterations; each iteration includes a call to the QPU for each value of $h$ probed.

\subsection*{Graph embedding}

The qubits in the QA processor are intercoupled in a ``Pegasus'' layout \cite{Boothby2020}, in which a qubit is coupled to up to 15 other qubits.  From these available couplers we select a geometry that represents a kagome graph using three qubits per kagome spin as depicted in Fig.~1c.  We show the full embedded lattice in Fig. \ref{fig:embedding}.  The  kagome embedding does not require the use of all qubits, and it is possible to embed a defect-free lattice with no site vacancies, despite the existence of some inoperable qubits (empty circle in Fig.~\ref{fig:embedding}c).

Since ferromagnetic chains are sometimes broken, they are majority-voted to provide an unambiguous mapping from the qubit system to the kagome system.  We run all experiments presented herein with $J_{ij}=0.9$ for AFM couplers and $J_{ij}=-1.5$ for FM couplers.  This choice of ferromagnetic coupling is sufficient to guarantee that chains are almost never broken in QPU output, despite the frustration in the system.

\subsection*{Disorder suppression}

In this application we perform many experiments on a single programmed lattice, whose classical ground state is highly degenerate.  Under such conditions it is appropriate to refine the general-purpose QA calibration by exploiting symmetries in the system.

For example, when $h=0$ each qubit should have average magnetization $\braket{s_i}=0$.  Thus we tune per-qubit flux offsets to balance qubits at zero for $h=0$, then use the same flux offsets when $h\neq 0$.  In this experiment, we are not interested in probing boundary conditions.  Rather, we want to simulate the thermodynamic limit of an infinite system.  In an infinite system, for any fixed $h$, the correlation of two neighboring kagome sites $\braket{s_is_j}$ is the same.  Thus we fine-tune the AFM couplers to promote this property.  Since the three-qubit FM chains are almost never broken, we do not fine-tune the FM couplers.  Similarly, for any fixed $h\neq 0$, the magnetization of each qubit should be equal; we fine-tune the per-qubit fields $h_i$ to promote this property (maintaining the property that the average $\tfrac 1 N\sum_{i}h_i$ does not change from the nominal value $h$).  These calibration refinements are performed before collecting the analyzed data.  Fig.~\ref{fig:shim} shows an example of this refinement for $J=0.9$, $h=0.6$, with the magnetizations and correlations achieved, and the programmed values that achieve them.

\clearpage

\begin{figure*}[htp]
  \includegraphics[width=14cm]{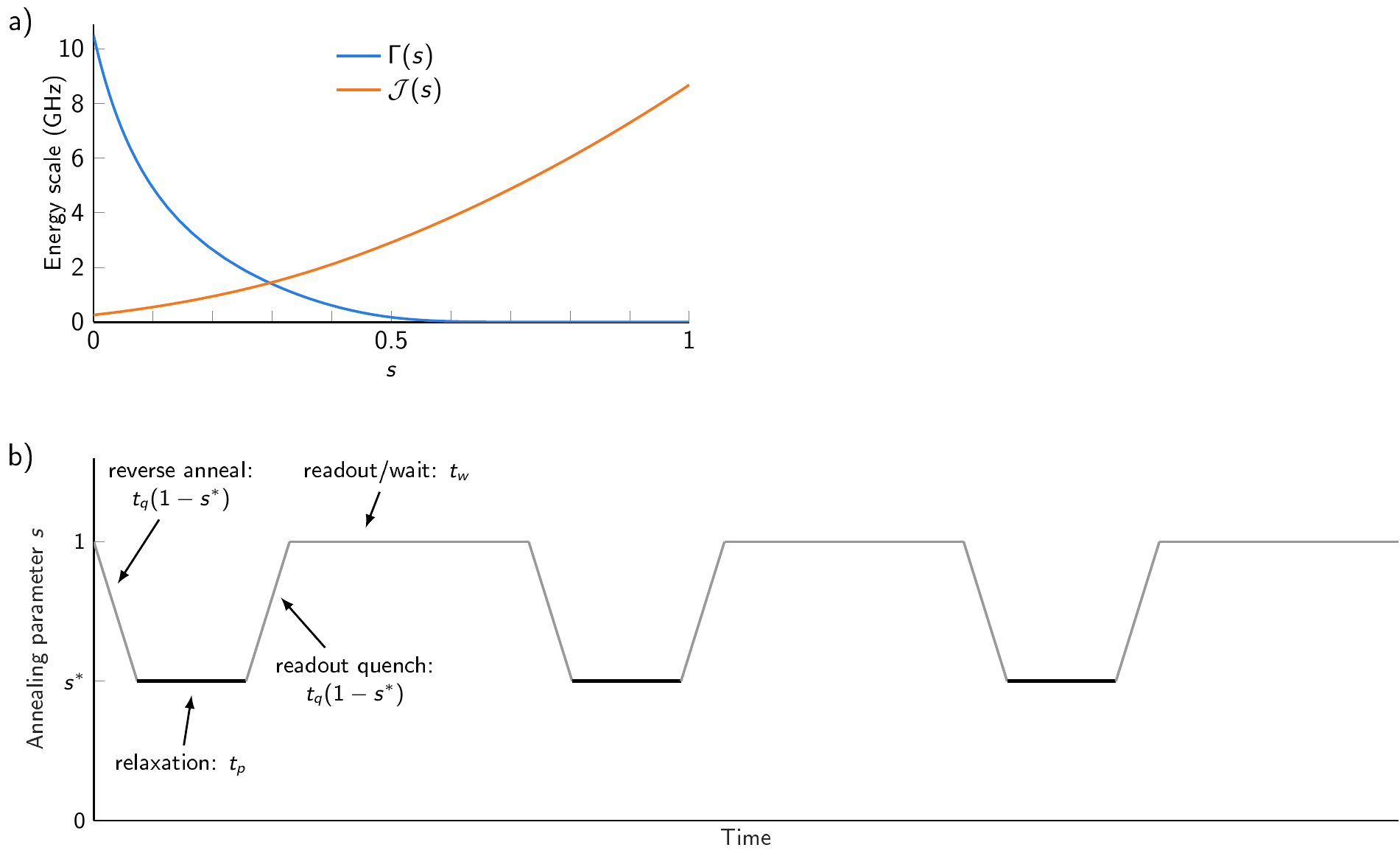}
  \caption{{\bf Quantum annealing schedule and protocol.} {\bf a}, Transverse field $\Gamma(s)$ and Ising energy scale $\mathcal J(s)$ as a function of annealing parameter $s$.  Note that the total coupling between two three-qubit chains is $1.8\mathcal J$.  {\bf b}, Quantum evolution Monte Carlo method.  A sequence of classical readout states is generated by repeated exposure to quantum fluctuations and thermal fluctuations.}
  \label{fig:protocol}\end{figure*}
\clearpage

\begin{figure*}[htp]
  \includegraphics[width=\linewidth]{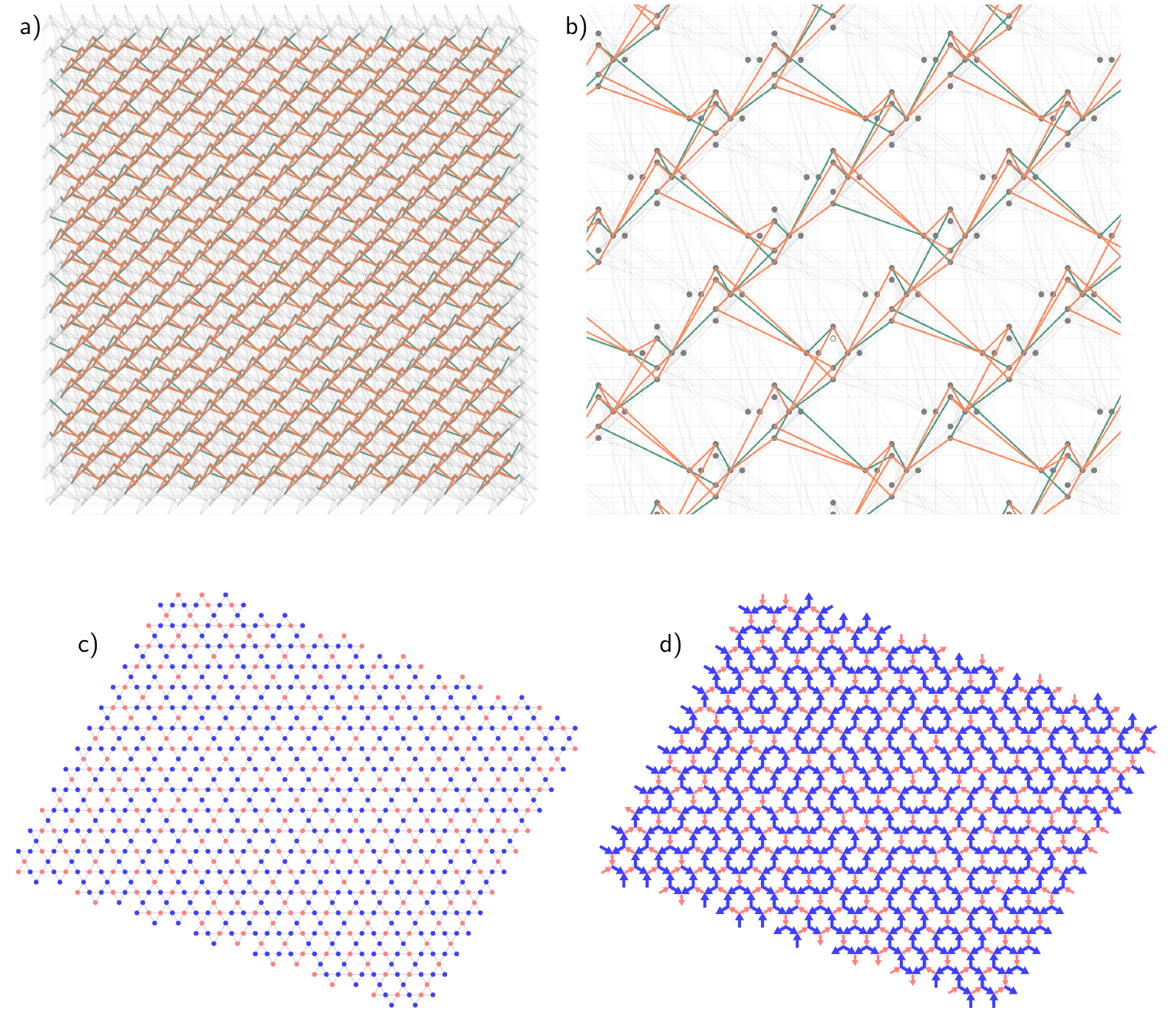}
  \caption{{\bf Embedding of the kagome lattice into the qubit graph.} {\bf a--b}, Each kagome site is represented by three qubits, coupled together ferromagnetically in a chain ($J_{ij}=-1.5$).  The entire qubit graph and embedding are shown in {\bf a} with green and orange lines representing FM and AFM couplers respectively; {\bf b} shows a detailed zoom, with operable and inoperable qubits represented by filled and empty circles respectively.  {\bf c--d}, The embedding shown in {\bf a} realizes a 729-site kagome lattice, which can be viewed as Ising spins ({\bf c}), or magnetic dipoles ({\bf d}).} 
\label{fig:embedding}\end{figure*}
\clearpage

\begin{figure*}[htp]
  \includegraphics[width=.8\linewidth]{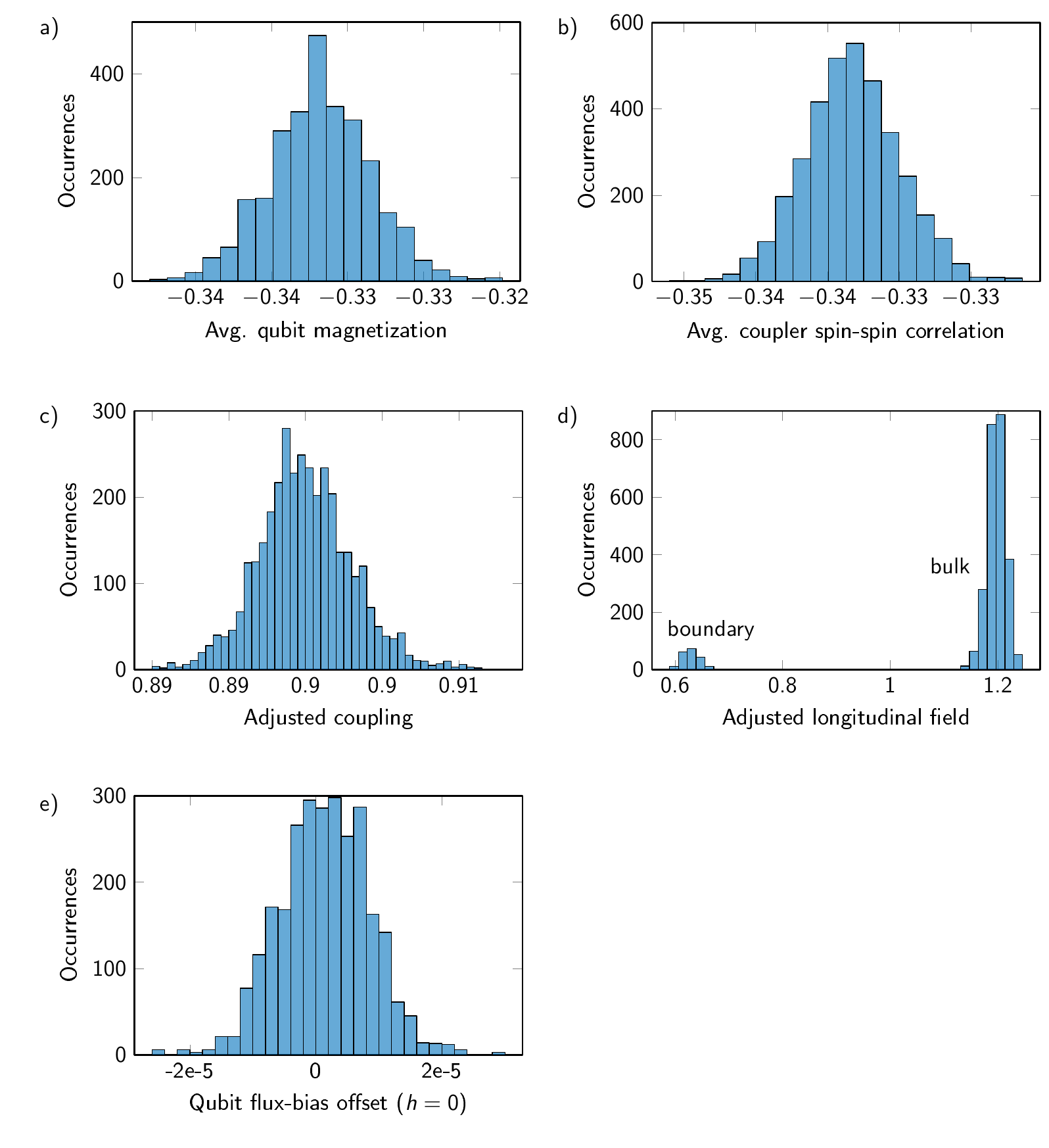}
  \caption{{\bf Suppressed disorder with fine-tuned Hamiltonian terms.} Example data are shown for nominal AFM coupling values of $J=0.9$ and local fields of $h=0.6$.  {\bf a--b,} Over 100 iterations, tightly-concentrated average qubit magnetizations and spin-spin correlations of coupled pairs indicate a balanced degenerate ice system.  {\bf c--e}, This is achieved by small adjustements of couplers ({\bf c}), adjustment of fields ({\bf d}), and qubits are balanced using flux-bias offsets at $h=0$ that are also used at nonzero $h$.  Note the two modes in {\bf d}, where boundary spins are assigned roughly half the field of bulk spins, in accordance with their degree in the graph, to achieve similar magnetizations.} 
\label{fig:shim}\end{figure*}
\clearpage

\end{document}